\documentclass[backref, breaklinks, twocolumn, colorlinks]{aastex631}   
\hypersetup{linkcolor=blue,citecolor=blue,filecolor=cyan,urlcolor=black}
 
\usepackage{latexsym,amsmath,amssymb}
\usepackage{verbatim}
\usepackage{booktabs}
\usepackage{multirow}
\usepackage{color}
\usepackage[hang]{footmisc}
\usepackage{soul,xcolor}
\usepackage{afterpage}
\usepackage{cuted}
\usepackage{tabularx}

\bibpunct{(}{)}{;}{a}{}{,}

\newcommand {\chandra} {\textsl{Chandra}}
\newcommand {\swift} {\textsl{Swift}}

\newcommand {\fermi} {\textsl{Fermi}}
\newcommand {\nustar} {\textsl{NuSTAR}}

\newcommand {\nicer} {\textsl{NICER}}
\newcommand {\ixpe} {\textsl{IXPE}}

\def\deg{$^\circ$}
\def \rsun {\ifmmode$R$_{\odot}\else R$_{\odot}$}

          \font\sixrm=cmr6

\def\sigt{\sigma_{\hbox{\sixrm T}}}
\def\taut{\tau_{\hbox{\sixrm T}}}

\def\rns{R_{\hbox{\sixrm NS}}}

\def\teq#1{$\, #1\,$}                         

\def \hcm {\hbox {\ifmmode $ atoms cm$^{-2}\else atoms cm$^{-2}$\fi}}

\def\approxgt{\mathrel{\hbox{\rlap{\lower.55ex \hbox {$\sim$}}
        \kern-.3em \raise.4ex \hbox{$>$}}}}
\def\approxlt{\mathrel{\hbox{\rlap{\lower.55ex \hbox {$\sim$}}
        \kern-.3em \raise.4ex \hbox{$<$}}}}

\def \arcmin {\hbox{$^\prime$}}
\def \arcsec {\hbox{$^{\prime\prime}$}}

\newcommand\T{\rule{0pt}{2.6ex}}
\newcommand\B{\rule[-1.2ex]{0pt}{0pt}}

\def \src {1E~1841$-$045}

\begin{document}
\setstcolor{red}

\title{\rm \uppercase{X-ray polarization of the magnetar \src}}

\author[0000-0002-0254-5915]{Rachael~Stewart}
\affiliation{Department of Physics, The George Washington University, Washington, DC 20052, USA}

\author[0000-0002-7991-028X]{George~A.~Younes}
\affiliation{Astrophysics Science Division, NASA Goddard Space Flight Center, Greenbelt, MD 20771, USA}
\affiliation{Center for Space Sciences and Technology, University of Maryland, Baltimore County, Baltimore, MD 21250, USA}

\author[0000-0001-6119-859X]{Alice~K.~Harding}
\affiliation{Theoretical Division, Los Alamos National Laboratory, Los Alamos, NM 87545, USA}

\author[0000-0002-9249-0515]{Zorawar Wadiasingh}
\affiliation{Department of Astronomy, University of Maryland, College Park, Maryland 20742, USA}
\affiliation{Astrophysics Science Division, NASA Goddard Space Flight Center,Greenbelt, MD 20771, USA}
\affiliation{Center for Research and Exploration in Space Science and Technology, NASA/GSFC, Greenbelt, Maryland 20771, USA}

\author[0000-0003-4433-1365]{Matthew~G.~Baring}
\affiliation{Department of Physics and Astronomy - MS 108, Rice University, 6100 Main Street, Houston, Texas 77251-1892, USA}

\author[0000-0002-6548-5622]{Michela Negro}
\affiliation{Department of Physics \& Astronomy, Louisiana State University, Baton Rouge, LA 70803, USA}

\author[0000-0001-7681-5845]{Tod E. Strohmayer} 
\affiliation{Astrophysics Science Division, NASA Goddard Space Flight Center, Greenbelt, MD 20771, USA}

\author[0000-0002-6089-6836]{Wynn~C.~G.~Ho}
\affiliation{Department of Physics and Astronomy, Haverford College, 370 Lancaster Avenue, Haverford, PA 19041, USA}

\author[0000-0002-0940-6563]{Mason Ng}
\affiliation{Department of Physics, McGill University, 3600 rue University, Montr\'{e}al, QC H3A 2T8, Canada}
\affiliation{Trottier Space Institute, McGill University, 3550 rue University, Montr\'{e}al, QC H3A 2A7, Canada}

\author{Zaven Arzoumanian}
\affiliation{Astrophysics Science Division, NASA Goddard Space Flight Center, Greenbelt, Maryland 20771, USA}

\author[0000-0001-9268-5577]{Hoa Dinh Thi}
\affiliation{Department of Physics and Astronomy - MS 108, Rice University, 6100 Main Street, Houston, Texas 77251-1892, USA}

\author[0000-0002-7574-1298]{Niccol\`o Di Lalla}
\affiliation{W. W. Hansen Experimental Physics Laboratory, Kavli Institute for Particle Astrophysics and Cosmology, Department of Physics and SLAC National Accelerator Laboratory, Stanford University, Stanford, CA 94305, USA}

\author[0000-0003-1244-3100]{Teruaki Enoto}
\affiliation{RIKEN Cluster for Pioneering Research, 2-1 Hirosawa, Wako, Saitama 351-0198, Japan}

\author[0000-0001-7115-2819]{Keith Gendreau}
\affiliation{Astrophysics Science Division, NASA Goddard Space Flight Center, Greenbelt, Maryland 20771, USA}

\author[0000-0001-8551-2002]{Chin-Ping Hu}
\affiliation{Department of Physics, National Changhua University of Education, Changhua 50007, Taiwan}

\author[0000-0002-3905-4853]{Alex Van Kooten}
\affiliation{Department of Physics, The George Washington University, Washington, DC 20052, USA}

\author[0000-0003-1443-593X]{Chryssa~Kouveliotou}
\affiliation{Department of Physics, The George Washington University, Washington, DC 20052, USA}

\author{Alexander McEwen}
\affiliation{Center for Space Sciences and Technology, University of Maryland, Baltimore County, Baltimore, MD 21250, USA}

\begin{abstract}
We report on IXPE and NuSTAR observations beginning forty days after the 2024 outburst onset of magnetar 1E 1841-045, marking the first IXPE observation of a magnetar in an enhanced state. Our spectropolarimetric analysis indicates that both a blackbody (BB) plus double power-law (PL) and a double blackbody plus power-law spectral model fit the phase-averaged intensity data well, with a hard PL tail ($\Gamma=1.19$ and 1.35, respectively) dominating above $\approx5$ keV. For the former model, we find the soft PL (the dominant component at soft energies) exhibits a polarization degree (PD) of $\approx30\%$ while the hard PL displays a PD of $\approx40\%$. Similarly, the cool BB of the 2BB+PL model possesses a PD of $\approx15\%$ and a hard PL PD of $\approx57\%$. For both models, each component has a polarization angle (PA) compatible with celestial north. Model-independent polarization analysis supports these results, wherein the PD increases from $\approx15\%$ to $\approx70\%$ in the 2--3~keV and 6--8~keV ranges, respectively, while the PA remains nearly constant. We find marginal evidence for phase-dependent variability of the polarization properties, namely a higher PD at phases coinciding with the hard X-ray pulse peak. We compare the hard X-ray PL to the expectation from resonant inverse Compton scattering (RICS) and secondary pair cascade synchrotron radiation from primary high-energy RICS photons; both present reasonable spectropolarimetric agreement with the data, albeit, the latter more naturally. We suggest that the soft PL X-ray component may originate from a Comptonized corona in the inner magnetosphere.

\end{abstract}



\section{Introduction} \label{sec:intro}

Magnetars are endowed with the strongest magnetic fields among the neutron star population, often surpassing $10^{14}$~G \citep{kouveliotou98Nat:1806}. This magnetic energy reservoir is evident through the magnetar's bright persistent X-ray emission, unique bursting abilities, and outburst epochs spanning months to years \citep[see, e.g.,][for a review]{turolla15:mag, kaspi17:magnetars}. Additionally, due to the presence of such large B-fields, the X-ray radiation emanating from magnetars is expected to be highly polarized \citep{1974A&A....36..379G,1988ApJ...324.1056M,Heyl-2000-MNRAS,Heyl-2002-PRD,2009MNRAS.399.1523V,taverna20MNRAS,Caiazzo-2022-MNRAS}, either in the ordinary (O) mode or the extraordinary mode (X), parallel or perpendicular to the plane of the magnetic field and photon wave vector, respectively. Exotic quantum electrodynamics (QED) effects in magnetar regimes are also expected to play a significant role in shaping their polarization characteristics, such as vacuum birefringence \citep{1992herm.book.....M,harding06RPPh} and mode conversion \citep{Lai-Ho-2003PhRvL,2006MNRAS.373.1495V,2017ApJ...850..185Y,lai23PNAS}, among others. Lastly, such polarization properties are highly energy- and phase-dependent, and a function of surface properties (gaseous versus condensate) and composition \citep[see, e.g.,][for a review]{taverna24Galax}.

Magnetars in quiescence and in outburst typically exhibit multiple spectral components: a thermal blackbody-like component (generally below 2 keV), a soft power law and a hard power law (generally above 5 keV).  The blackbody (BB) component has been modeled as a magnetized atmosphere \citep[see, e.g.,][]{2006MNRAS.373.1495V} or as condensed surface emission \citep{vanAdelsberg2005,2012A&A...546A.121P}.  The soft power law (SPL) component may be due to Comptonization of atmosphere radiation by a corona just above the neutron star surface or resonant scattering by mildly relativistic electrons at higher altitude \citep{nobili08MNRAS:XraySpec}. The hard power law (HPL) component has been attributed to resonant inverse Compton scattering by relativistic electrons \citep{BH-2007-ApandSS} or to annihilation bremsstrahlung by a dense, trans-relativistic pair plasma \citep{Thompson2020}. While spectral and timing information alone are limited in their ability to discriminate between these models, their distinctive polarization characteristics, when confronted with observations, could provide critical information for their true contribution, which in turn help distinguish competing models.

The persistently bright magnetar \src, located at approximately 8.5 kpc \citep[][revised to $5.8\pm 0.3$ kpc by \citealt{ranasinghe_aj2018}]{tian08ApJkes73} in the center of the X-ray and radio supernova remnant (SNR) Kes 73, has an inferred surface dipolar magnetic field of about $6.9\times10^{14}$~G at the equator, one of the largest among the population \citep{vasisht97ApJ}. The source also exhibits a bright hard X-ray tail, which extends to $>200$~keV \citep[][see also \citealt{2017ApJS..231....8E, an13ApJ:1841, an15ApJ:1841}]{kuiper04ApJ}. Since its discovery, the source has displayed sporadic bursting activity on several occasions \citep{kumar10ApJ:1841, lin11ApJ:1841, dib14ApJ}. Yet, no enhanced emission to its quiescent flux has ever been confirmed following these bursting episodes \citep{lin11ApJ:1841, dib14ApJ}. On 2024 August 21, \swift's Burst Alert Telescope (\textit{BAT}) reported the detection of bursting activity from \src\ starting at 19:01:18 UT \citep{dichiara24ATel16784}, later corroborated with \fermi's Gamma-ray Burst Monitor (\textit{GBM}) and the Neutron star Interior Composition Explorer (\nicer) \citep{roberts24ATel16786, ng24ATel16789}, among other high energy instruments. \nicer, \swift's X-Ray Telescope (\textit{XRT}), and Nuclear Spectroscopic Telescope ARray (\nustar) observations 8 days following the initial report revealed that the source 2--70 keV flux increased by a factor 2 and that the soft and hard X-ray pulse shape noticeably changed to a multi-peaked structure \citep{younes24ATel16802}. These observations confirmed that the source is undergoing a radiative outburst in conjunction to its bursting activity, its first confirmed outburst.

In this paper, we report the polarimetric and spectropolarimetric results from the magnetar \src\ during its 2024 outburst epoch, through the analysis of a simultaneous Imaging X-ray Polarimetry Explorer (\ixpe) and \nustar\ observation, taken between September 28 and October 10, 40--50 days following its outburst onset \citep{dichiara24ATel16784}. At the time of the observation, the source flux was still 40\% elevated compared to its baseline emission, predominantly in the hard X-ray tail, and its pulse profile displayed evident variation compared to quiescence. This is the first time that \ixpe\ has observed a magnetar in an enhanced flux state, contrasting with all its previous magnetar observations \citep[e.g.,][]{taverna22Sci, lai23PNAS, zane23ApJ, turolla23ApJ:1806, heyl24MNRAS:2259}. Section~\ref{sec:obs} summarizes the observations and data reduction procedures. Section~\ref{sec:results} details our findings. Section~\ref{sec:disc} discusses our results in the context of radiation processes in high B-field regime. We note that a companion paper also details the analysis of the \ixpe\ and \nustar\ data while presenting a slightly different approach to the spectropolarimetry part (Rigoselli et al. 2025). The results and theoretical interpretation are largely consistent, except for the discussion on the nature of the soft X-ray component, where the two papers present distinct interpretive elements.

\section{Observations and data reduction}
\label{sec:obs}

The Imaging X-ray Polarimetry Explorer (\ixpe) consists of three identical telescopes, each equipped with a Detector Unit (DU) that houses a Gas Pixel Detector and is sensitive in the 2--8 keV energy range (\citealt{weisskopf22JATIS}; \citealt{baldini2021}). \ixpe\ observed the magnetar \src\ for 170~ks from 2024 September 28 to October 1 and for 130~ks from October 7 to October 10. The total livetime for the observation is about 292.5~ks. 

We perform the data cleaning, region selection, and analysis utilizing the software \texttt{ixpeobssim} v.31.0.1 \citep{baldini_2022_softX_ixpe}. We set the instrument response functions (IRFs) to \texttt{ixpe:obssim20240701:v13} from \texttt{ixpeobssim}'s pseudo-CALDB which contains the arf, rmf, PSF, vignetting functions, modulation factors, and modulation responses corresponding to the observation date \citep{baldini_2022_softX_ixpe}. We applied the solar-system barycenter correction utilizing the HEASoft tool \texttt{barycorr} and best-known source coordinates \citep[280.3275732\deg, -4.9339910\deg;][]{wachter04ApJ:1841}.

We derive the rotational phase of each photon using  \texttt{CRIMP}\footnote{\url{https://github.com/georgeyounes/CRIMP/tree/main}} (Code for Rotational-analysis of Isolated Magnetars and Pulsars) and a timing solution covering the entirety of 2024\citep{younes2025arxiv}. The \texttt{ixpeobssim} task \texttt{xpselect} was used to specify the energy and phase bins used in creating the energy- and phase-resolved data products, e.g., the event files and pulse profiles. Lastly, we use \texttt{xpbin} to extract all binned products, including the pulse profiles, polarization cubes, and PHA1, PHA1Q, and PHA1U spectra.  For spectropolarimetry, we utilize Xspec version 12.14.0c, part of HEASOFT 6.33.1, to perform the spectral fitting.

We generate and examine radial profiles in order to select extraction region sizes that suitably maximize the signal-to-noise ratio of the source while minimizing the contribution from the surrounding SNR. We select a 30\arcsec\ circular region for the source and an annulus of 60\arcsec and 120\arcsec\ for the background inner and outer radii, respectively, centered on the brightest pixel. The latter encompasses emission predominantly from the SNR, in addition to the cosmic background. Its area-normalized flux contributes 12\%, 5\%, and $<$2\%  to the point source flux at the 2--3 keV, 3--4 keV, 4--8 keV ranges, respectively. The modest elevated contribution at lower energies is due to the SNR emission. Moreover, the normalized Stokes parameters \textit{Q/I} and \textit{U/I} show no significant polarization signal from the SNR\footnote{We are referring to the linear polarization in reference to the celestial north direction. We tested the presence of radial and tangential polarization in the SNR region finding values below the MDP$_{99}\sim$8\%.} with the measured PD falling under
the MDP$_{99}\sim$13\%. Hence, we deem that the contamination from the SNR to the magnetar emission is negligible, and that subtraction of the SNR + cosmic contribution from the polarization cubes and spectra is a sufficient treatment of the background for the remainder of this analysis.

We performed a search for short bursts in the \ixpe\ data. We identify one candidate at 60581.431488977745 MJD (TDB), with a total number of 8 counts in a 22 ms bin. This is highly unlikely to occur by chance according to a Poisson probability density given the source rate of $\sim 0.1$ cts/s. We eliminate this time interval in our analysis. 
The background rejection outlined by \cite{DiMarcoCut} disproportionately cut photons compatible with the location of the magnetar, especially at the higher energies between 6 and 8 keV, rejecting up to 10\% of the source photons. On the other hand, at low energies, the background rejection does not improve the polarization detection significantly. We therefore opted to not apply the background rejection for this source (see the Appendix for more details). We similarly abstained from performing solar deflaring since the rejection did not improve the S/N but sacrificed a non-trivial number of counts.

The Nuclear Spectroscopic Telescope ARray (\nustar;\, \citealt{harrison13ApJ:NuSTAR}) observed \src\ on 2024 September 28 with a 50 ks exposure (observation ID 91001335002), simultaneous with the first leg of the \ixpe\ observation. \nustar\ consists of two identical focal plane modules, FPMA and FPMB. We utilize {\tt NuSTARDAS} software version 2.1.2 to clean and calibrate the event files. We employed the task {\tt nuproducts} to extract source events, light curves, and spectra from a circular region with a 30\arcsec-radius around the source central brightest pixel. For the background region, we extract an annulus centered around the source with inner and outer radii of 60\arcsec and 120\arcsec, respectively, which encompasses part of the SNR contribution and cosmic background.


The Neutron star Interior Composition Explorer (\nicer) is a non-imaging, soft X-ray telescope mounted on the International Space Station, with a $\approx6\arcmin$ diameter field-of-view (FoV; \citealt{gendreau16SPIE}), and with a collecting area peaking at $1900$~cm$^2$ at 1.5~keV and a timing resolution $<300$\,ns. We cleaned and calibrated all \nicer\ data utilizing {\tt NICERDAS} version v12 as part of {\tt HEASoft} version 6.33. We applied the standard filtering criteria utilizing the task {\tt nicerl2}. We compared light curves (binned with a 5~second resolution) created in the energy range 2--8~keV (where the source is expected to dominate the emission) to ones created in the 12--15~keV energy range. Almost all the counts in the latter band are due to high energy particle background. We eliminated any simultaneous flaring background intervals that appeared in both light curves. For this paper, we only present the pre-outburst \nicer\ data covering the February to August 2024 period for comparison purposes with the post-outburst simultaneous \ixpe\ and \nustar\ data.

\section{Results}
\label{sec:results}
\subsection{Polarization Characteristics}
\begin{figure}[t!]
    \centering
    \includegraphics[width=\linewidth]{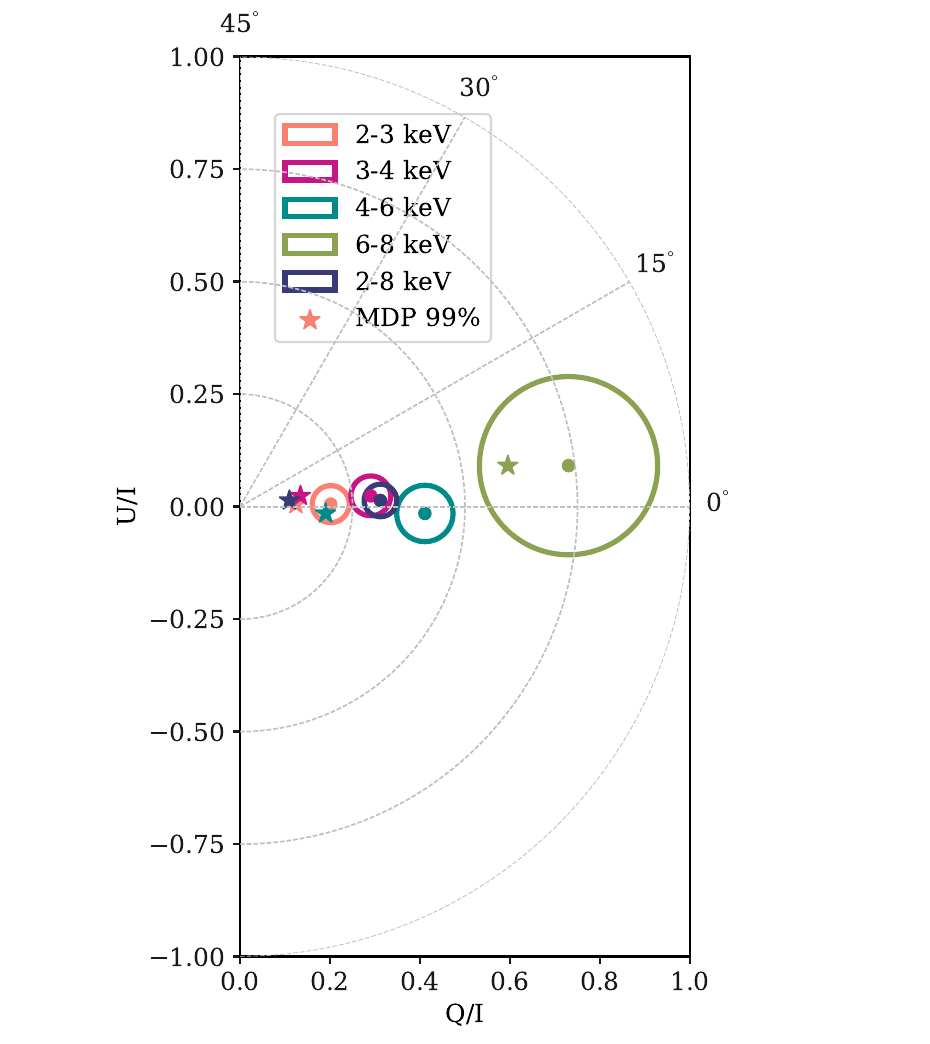}
    \caption{The background-subtracted, phase-averaged polarization characteristics of \src\ according to 4 energy bins. Contours show the $68.3\%$ ($1 \sigma$) confidence regions for both polarization degree and angle. Stars show the corresponding MDP$_{99}$ for each energy band. The energy-integrated polarization is shown in dark blue.}
    \label{fig:energy_resolved_polarization}
\end{figure}

Figure \ref{fig:energy_resolved_polarization} shows the phase-averaged \textit{Q} and \textit{U} stokes parameters from the combined DUs measured in different energy bands with the circle marking the 68.3\% confidence level. The phase- and energy-integrated polarization angle (PA) is $1\pm 3$\deg~east of north, and the polarization degree (PD) from $2-8$ keV is $31\pm 4\%$. This measured polarization degree is large, comparable to that of 1RXS J170849.0$-$400910, which has the strongest measured phase- and energy-averaged polarization signal among the magnetar population to date (\citealt{zane23ApJ}, \citealt{taverna24Galax}). 

We find higher polarization degrees by binning the data according to four energy bands, i.e. 2--3, 3--4, 4--6, and 6--8 keV. This energy-resolved analysis reveals that the polarization degree increases monotonically according to energy, from $20\pm 4\%$ at 2--3 keV up to $\approx 75\pm 
20\%$ at 6--8 keV (Figure~\ref{fig:energy_resolved_polarization}). Meanwhile, the PA remains consistent with $0$\deg~ across all energy bands. We note that all energy-dependent polarization values surpass the 99\% C.L. minimum detectable polarization (MDP$_{99}$) at the corresponding energies.

\begin{figure}[t!]
    \centering
    \includegraphics[width=\linewidth]{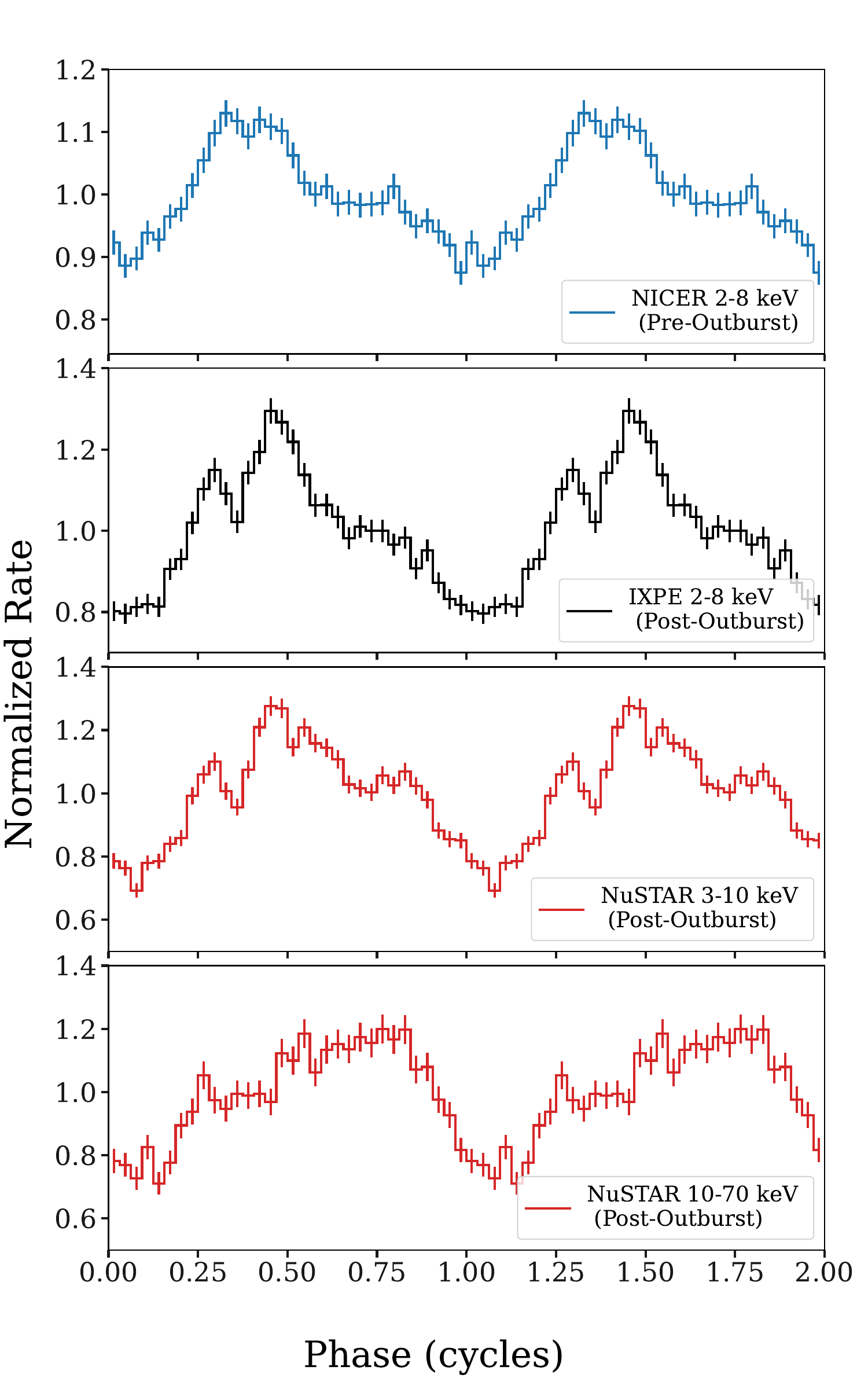}
    \caption{\textit{Upper Panel}: The normalized intensity pulse profile of \src\ from $2-8$ keV using observations from \nicer~ preceding the outburst. \textit{Second Panel}: The source 2-8~keV intensity pulse profile post-outburst as observed with \ixpe. \textit{Third Panel}: The \nustar\ pulse profile from $3-10$ keV simultaneous to \ixpe. \textit{Bottom Panel}: The \nustar\ 10-70~keV pulse profile.}
    \label{fig:NICER_IXPE_PP}
\end{figure}

Figure \ref{fig:NICER_IXPE_PP} shows the \ixpe+\nustar\ post-outburst pulse profiles (second, third, and fourth panels). The pulse profile of \src\ preceding the outburst as observed with \nicer\ is also displayed (upper panel). We fold the events using a phase-connected timing solution derived from a long-term \nicer\ monitoring campaign according to a frequency of $\nu = 0.084702008(1)$ Hz and a frequency derivative of $\dot{\nu} =  -2.938(2)\times 10^{-13}$ Hz s$^{-1}$ at an epoch of 60450 MJD TDB (see \citealt{younes2025arxiv} for full description of timing solution, including treatment of a glitch observed at epoch 60543 MJD). Each pulse profile uses 32 bins and is normalized by its average. Two cycles are shown for clarity. The pulse profile morphology of \src\ clearly evolves following the outburst. The soft X-ray (e.g. 2--8, 3--10 keV) main pulse peak narrows in width while a subpeak emerges on its leading edge. The hard X-ray 10--70 keV post-outburst pulse profile, while broad, peaks at a phase coincident with the shoulder observed at lower energies. This pulse shape is slightly different when compared to historical \nustar\ profiles which mainly show a flat-top pulse \citep[][and  \citealt{younes2025arxiv}]{an13ApJ:1841, an15ApJ:1841}.
   
\begin{figure*}[t!]
    \centering
    \includegraphics[width=\textwidth]{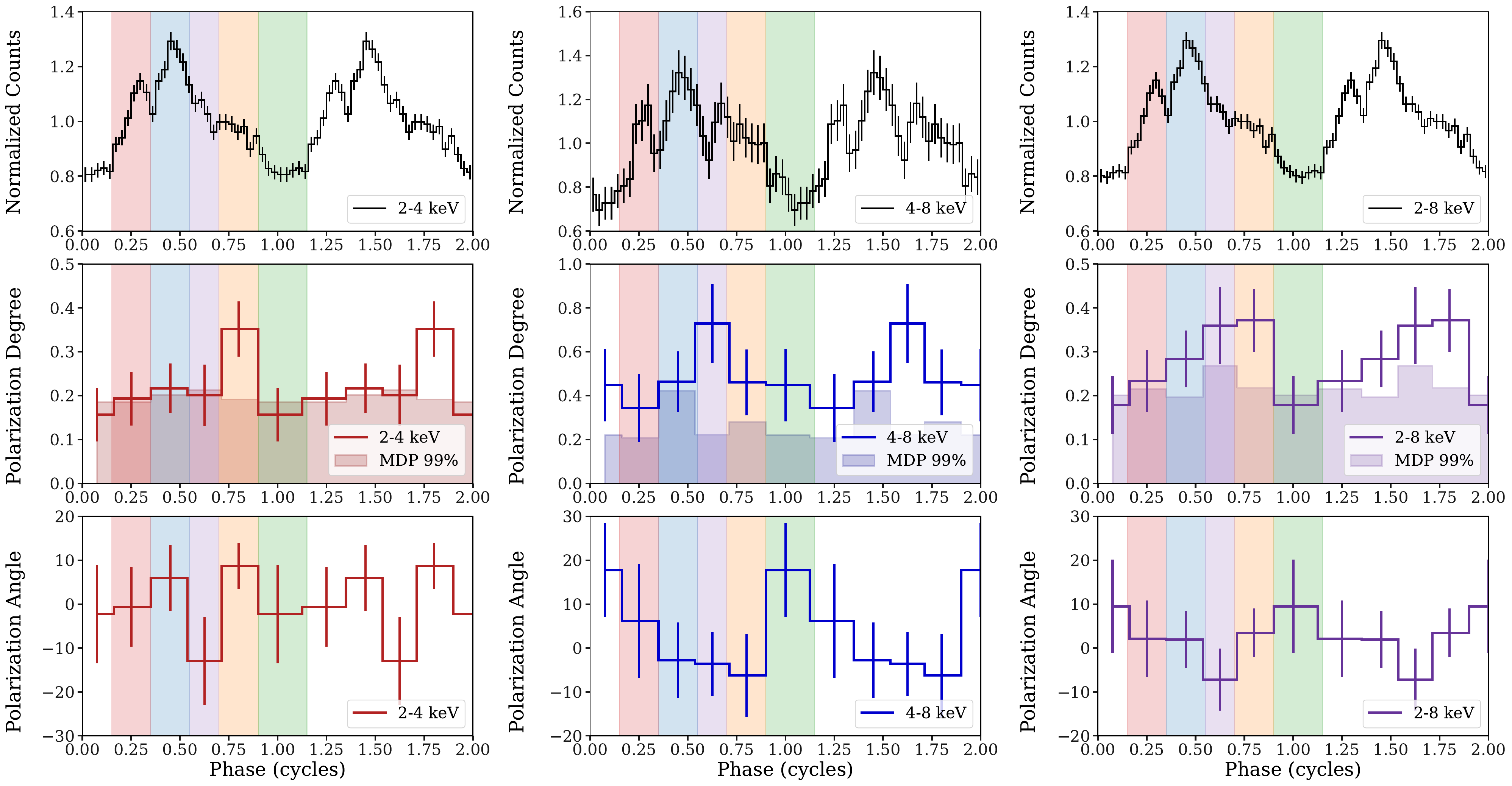}
    \caption{Phase-resolved polarization characteristics of \src. The columns display the intensity pulse profiles (top row), polarization degree (center), and polarization angle (bottom row) according to 2--4 keV (left column), 4--8 keV (central column), and 2--8 keV (right column). The phaseograms are divided according to five phase bins (shown in colored vertical bands) that correspond to morphological features of the \ixpe\ energy-integrated intensity pulse profile. The corresponding MDP$_{99}$ for each phase and energy bin are shown as overlapping shaded regions.}
    \label{fig:phase_resolved_pol}
\end{figure*}

The upper panels of Figure~\ref{fig:phase_resolved_pol} shows the \ixpe\ pulse profiles in three energy bands 2--4, 4--8, and 2--8 keV. We present the corresponding phase-resolved polarization degree and angle in a select five phase bins in the middle and lower panels, respectively. These phase-bins were chosen to correspond to the morphological features of the pulse, namely, the newly-formed subpeak (red-shaded region), main pulse (blue-shaded region), shoulder (purple- and orange-shaded regions) and the pulse minimum (green-shaded region). We note that the purple and orange-shaded regions coincide with the pulse-maximum of the hard 10--70 keV X-ray band. The horizontal shaded areas in the polarization degree panels display the MDP$_{99}$ for each phase bin at each corresponding energy.

The phase-resolved polarization degree of the 2--8 keV band surpasses the MDP$_{99}$ at each bin with the exception of the pulse minimum (in green). The polarization degree exhibits small variability between ~$20-40\%,$ with the PD minimum aligned with the intensity pulse profile minimum and the maximum with the trailing shoulder (shown in purple and orange), which corresponds to the hard X-ray peak. Only mild variation is detected in the polarization angle, with the value remaining largely consistent with 0\deg~ at each phase bin.

In the 2--4 keV range, only the phase bin corresponding to the orange band exceeds the MDP$_{99}$. Therefore, the statistics are not robust enough to assert a definitive phase-dependence for the 2--4 keV band. The polarization angle at each phase bin also maintains around 0\deg.

The 4--8 keV polarization degree shows mild variation between $\sim 30\%$ and $60\%$ but is primarily consistent within uncertainties. The 4--8 keV PD peak coincides with the peak of the \nustar~ 10--70 keV pulse profile (see Figure \ref{fig:NICER_IXPE_PP}). The polarization angle varies from around 0\deg~ up  to 30\deg~ at the intensity pulse minimum, albeit with large uncertainties.

\subsection{Spectropolarimetry}

To gain insights into the spectral behavior of the polarization characteristics, we used Xspec \citep{arnaud96conf} to simultaneously fit the \nustar~ spectra in the 3--70 keV and the \ixpe\ spectra in the 2--8 keV range. For the \nustar\ observations, we grouped each spectra to have 5 counts per bin and used the W-statistic (\texttt{cstat} in Xspec); likewise for the \ixpe\ data, we grouped the spectra according to 25 counts per bin and used chi-squared statistics. Due to the \nustar\ and \ixpe's limited sensitivity to the neutral hydrogen column density in the direction of the source, we fix $N_{\text{H}}$ to the historical value of $2.6 \times 10^{22}$ cm$^{-2}$ as derived with \chandra\ \citep{kumar10ApJ:1841}. We likewise use the Anders \& Grevesse elemental abundances (\texttt{angr} in Xspec, \citealt{anders89gca}) for consistency along with the photo-electric cross-sections of \cite{verner96ApJ:crossSect} and the Tuebingen-Boulder ISM absorption model (\texttt{tbabs}). Lastly, we add a multiplicative constant component to the spectral models to take into account any calibration uncertainty between the \nustar\ FPM detectors and the \ixpe\ DUs. This constant also encompasses the difference in the SNR contribution to the \ixpe\ data, in which it can be largely resolved, compared to \nustar\, where it is blended with the source point spread function. We find the calibration constants for \ixpe\ to be about 70\% that of \nustar\ for each spectral model tested (see Table \ref{tab:model_params}).

\begin{figure*}[t!]
    \centering
    \includegraphics[width=\linewidth]{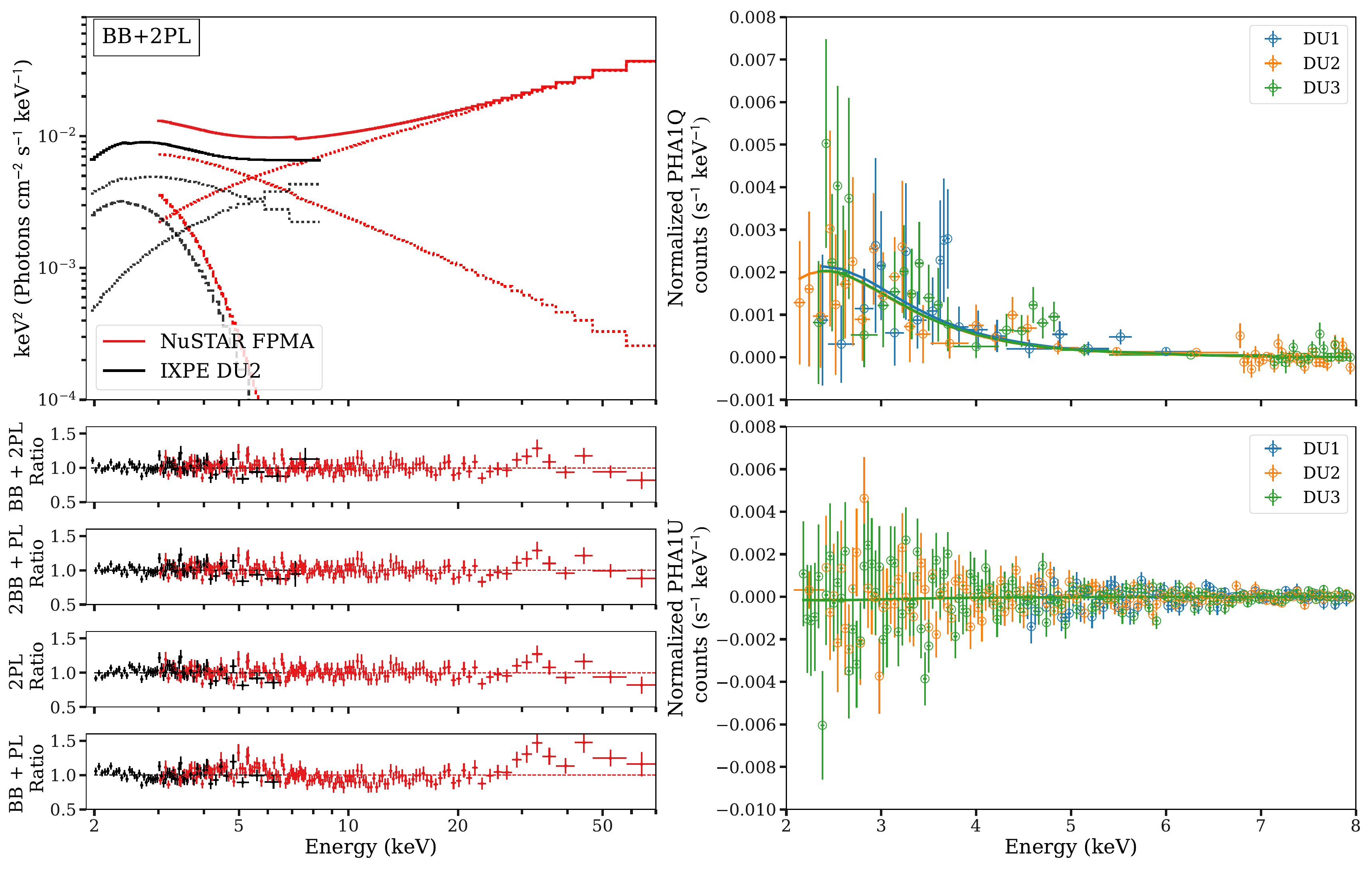}
    \caption{The joint spectral $\nu F_{\nu}$ models of the simultaneous \nustar+\ixpe\ observation. The upper left panel shows the best fit model of \texttt{constant*tbabs(polconst*bbodyrad+polconst*powerlaw+polconst*powerlaw)}. Note: Only \nustar~ FPMA and \textit{DU2} are displayed to enhance visual clarity. The second left panel shows the data divided by the folded model for the absorbed BB+2PL model; the third, fourth and bottom left panels show the ratio between the data and the folded model for an absorbed 2BB+PL, 2PL, and BB+PL for comparison. The right panels show the normalized Stokes \textit{Q} (top) and \textit{U} (bottom) spectra for the 3 \textit{IXPE} DUs with the solid lines showing the best fit model in linear space. Best-fit spectral parameters can be found in Table \ref{tab:model_params}.}
    \label{fig:phase_averaged_spectra}
\end{figure*}

We fit the \nustar+\ixpe\ intensity spectral data to a blackbody with double power-law (BB+2PL) model. We find that this model results in a statistically acceptable fit with a total fit statistic of 2001 for 2136 degrees of freedom (d.o.f). The best-fit $\nu F_{\nu}$ models of each component are shown in the upper left panel of Figure \ref{fig:phase_averaged_spectra}. The corresponding data-to-model ratio for the BB+2PL model is shown in the second left panel. We investigated an alternative double blackbody + powerlaw (2BB+PL) model, which performed comparably well to the BB+2PL with a total fit statistic of 2008 for 2136 d.o.f. (see Fig \ref{fig:phase_averaged_spectra} center left panel).

We found that a three component spectral model produced a slightly favorable fit over a 2PL model, which resulted in a total fit statistic of 2037 for 2138 d.o.f. A BB+PL model resulted in a considerably worse fit with a total fit statistics of 2197 for 2138 d.o.f. The latter resulted in strong systematic deviation from unity in the data-to-model ratio residuals as shown in the bottom left panel of Figure \ref{fig:phase_averaged_spectra}.


To probe the polarization properties of the BB+2PL best-fit spectral components, we fit the \textit{Q} and \textit{U} spectra alongside the \nustar+\ixpe\ intensity spectra utilizing a constant polarization model (\texttt{polconst} in Xspec). We show the phase-averaged \textit{Q} and \textit{U} Stokes spectra on the right-hand side of Figure \ref{fig:phase_averaged_spectra}.  We let all the intensity parameters freely vary, with the exception of $N_{\rm H}$ and obtain a BB temperature of kT$ = 0.42 \pm 0.01$, a soft PL photon index of $\Gamma_{\text{SPL}} = 3.1 \pm 0.2$ and a hard PL photon index of $\Gamma_{\text{HPL}} = 1.19^{+0.05}_{-0.06}$ (see Table \ref{tab:model_params} for the best-fit spectral parameters). The total unabsorbed 2--70 keV flux is $(1.03\pm 0.04) \times 10^{-11}$ erg~s$^{-1}$~cm$^{-2}$, approximately $30\%$ larger when compared to the historic value reported using \nustar\ \citep[][\citealt{younes2025arxiv}]{an13ApJ:1841}. The soft PL dominates above the BB emission by about a factor of 2.3 in the 2-8 keV range. At $\sim 5$ keV, the emission from the hard X-ray tail becomes dominant over the soft PL (Figure~\ref{fig:phase_averaged_spectra}). 

We cannot constrain the polarization properties of all three components simultaneously. We thus assume that the PD of one of the soft components is 0 (freezing the hard PL PD to 0 fails to constrain the fit). We find that upon freezing the soft PL PD to 0, the hard PL PD approaches 100\% polarization, which we deem physically unreasonable. We therefore freeze the BB PD to PD$_{\text{BB}}=0$, leading to a polarization degree of $30\pm10\%$ for the soft PL and $40\pm20\%$ for the hard PL. Additionally, the polarization angle lies within $1\sigma$ of 0\deg\ for both components. We also tested the impact of assigning the PD$_{\text{BB}}$ component a non-zero value in increments of 5\% up to 25\% for both models, based on the values presented in the model-independent polarization analysis (see Figure \ref{fig:energy_resolved_polarization}). We find that as PD$_{\text{BB}}$ increases, the SPL PD systematically decreases from a value of 30\% down to approximately 15\% while the hard PL increases from roughly 40\% to 55\%; notably, the PD and PA for each PL component remains within about 1$\sigma$ of the values corresponding to the PD$_{\text{BB}}=0$ case. This may be due to the fact that the flux of the BB is comparatively subdominant to that of the soft PL X-ray component as described above, and thus changing the assigned BB PD does not impact the fit much.

For the 2BB+PL model, we find a cool and a hot BB temperature of $0.41\pm0.01$~keV and $1.04\pm0.06$~keV, respectively, along with emitting areas of about $139\pm15$~km$^2$ and $0.58\pm0.14$~km$^2$ assuming a distance of 8.5 kpc. We find a photon index $\Gamma=1.34\pm0.02$. We again fail to constrain the fit of all three polarization components simultaneously. We therefore assume that the PD of one of the components is 0. If we assume the cool BB or the hard PL, which dominates at the lower and higher energy parts of the spectrum, respectively, to have ${\rm PD} = 0\%$, we find that the hot BB will exhibit a 100\% polarization in order to compensate for the lack of polarization signature at either end of the spectrum, which we consider nonphysical. Under the assumption of an unpolarized hot BB, we find that the cool BB and the PL have a PD of $15 \pm 6\%$ and $57 \pm 9\%$, respectively. The PA of both components are consistent with 0\deg\ at the $1\sigma$ level.
We again also examined the case in which the hot BB has a non-zero PD, and we find upon increasing PD$_{\text{BB}_{\text{hot}}}$ from 0\% to 25\% in increments of 5\%, the cool BB and hard PL polarization degree and angle fits do not vary beyond a few percent of those obtained with the PD$_{\text{BB}_{\text{hot}}}=0$ case.

We also investigated the spectro-polarimetric characteristics of the 2PL and BB+PL models, given that these fits still performed comparably well against the BB+2PL and 2BB+PL models (see Table \ref{tab:model_params}). For the 2PL model find a SPL PD of $0.17\pm0.06$, SPL PA of $(-3\pm 10)$\deg, HPL PD of $0.5\pm 0.1$, and a HPL PA of $(-1\pm 7)$\deg. For the BB+PL model we find a BB PD and PA of $0.3\pm0.1$ and $-3\pm9$\deg, respectively, and a HPL PD of $0.4\pm 0.2$ and HPL PA of $(-1\pm14)$\deg. Notably, the polarization characteristics obtained for the three-component spectral models with an unpolarized BB possess similar values as those from the two-component models.
\begin{table*}[t!]
\centering
\begin{tabular}{@{}ccccc@{}}
\toprule
Model Parameter & SPL + HPL  & BB + PL & BB + 2PL & 2BB + HPL\\ 
\midrule
$N_{\text{H}}$ (cm$^{-2}$)       & $2.6 \times 10^{22}$ (fixed) & ----- & ----- & ----- \\ 
C$_{\text{FPMA}}$              & $1.0$ (fixed)  & ----- & ----- & -----  \\
C$_{\text{FPMB}}$              & $1.04\pm 0.01$ & $1.04\pm 0.01$& $1.04\pm 0.01$  & $1.04\pm 0.01$ \\
C$_{\text{DU1}}$               & $0.73\pm 0.01$& $0.71\pm 0.01$& $0.71\pm 0.01$ & $0.70\pm 0.01$ \\
C$_{\text{DU2}}$               & $0.68\pm 0.01$ & $0.68\pm 0.01$& $0.67\pm 0.01$& $0.67\pm 0.01$ \\
C$_{\text{DU3}}$                & $0.70\pm 0.01$ & $0.70\pm 0.01$& $0.69\pm 0.01$& $0.69\pm 0.01$ \\
\hline
\multicolumn{5}{c}{Soft Component}\\
\hline
SPL Photon Index             & $3.84\pm 0.04$ & ----- & $3.1\pm0.2$ & ----- \\
PD$_{\text{SPL}}$        & $0.17\pm 0.06$ & ----- & $0.3\pm0.1$ & ----- \\
PA$_{\text{SPL}}$  & $(-3\pm 10)$\deg  & ----- & $(-3\pm9)$\deg & -----  \\

BB$_1$ kT (keV) & ----- & $0.467 \pm 0.005$ & $0.42 \pm 0.01$ & $1.04 \pm 0.06$ \\
PD$_{\text{BB}_1}$ & ----- & $0.12\pm 0.06$ & $0.0$ (fixed) & $0.0$ (fixed)\\
PA$_{\text{BB}_1}$  & ----- & $(-4\pm 16)$\deg & $0.0$ (fixed) & $0.0$ (fixed)  \\
BB$_2$  kT (keV)  & ----- & -----& -----& $0.41 \pm 0.01$ \\
PD$_{\text{BB}_2}$  & ----- & -----& ----- & $0.15\pm 0.06$ \\
PA$_{\text{BB}_2}$   & ----- & -----& ----- &  $(-4\pm 14)$\deg   \\
\hline
\multicolumn{5}{c}{Hard Component}\\
\hline
HPL Photon Index             & $1.23\pm0.02$ & $1.57\pm0.01$ & $1.19^{+0.05}_{-0.06}$ & $1.35\pm0.03$ \\
PD$_{\text{HPL}}$        & $0.5\pm0.1$  & $0.41\pm0.07$& $0.4\pm0.2$& $0.57\pm0.09$\\
PA$_{\text{HPL}}$  & $(-1\pm7)$\deg & $(-2 \pm 4)$\deg & $(-1 \pm 14)$\deg & $(-2 \pm 4)$\deg  \\

Total Fit Statistic               & 2037 with 2138 d.o.f. & 2197 with 2138 d.o.f. & 2001 with 2136 d.o.f. & 2008 with 2136 d.o.f.\\
\bottomrule
\end{tabular}
\caption{Spectral parameters comparing four models: an absorbed soft power-law (SPL) + hard power-law (HPL), an absorbed blackbody (BB) + power-law (PL) , an absorbed blackbody (BB) + soft power-law (SPL) + hard power-law (HPL), and an absorbed double blackbody + power-law (PL) obtained using Xspec. A multiplicative cross-calibration uncertainty was also added to the model with \nustar\ FPMA fixed to unity. All uncertainties are calculated at the 68.3\% confidence level. The fit statistic employs a combination of the C-stat and Chi-Squared statistics used for \nustar\ and \ixpe, respectively.}
\label{tab:model_params}
\end{table*}

\section{Discussion}
\label{sec:disc}

We have presented the detailed spectropolarimetric analysis of the first magnetar observed by \ixpe\ ( and \nustar) in an outburst, \src. These observations occurred while the source still exhibited a 30\% increase in flux mainly driven by the presence of an enhanced hard X-ray tail, which dominates the emission at energies $\gtrsim5$~keV, well within the \ixpe\ band. We explored various models to describe the spectral decomposition of the \ixpe+\nustar\ data. Both a BB+2PL and 2BB+PL models performed similarly well in capturing the broadband spectral shape. The spectro-polarimetric results for both fits are comparable, with the key distinction being the nature of the soft-energy spectral component as either thermal or non-thermal, albeit, both resulting in a comparable PD of about 15\% and 30\%, respectively. Additionally, both spectral models indicate that the hard X-ray tail is highly polarized with PD~$\approx40\%$ for the BB+2PL and PD~$\approx60\%$ for the 2BB+PL. The PA of both components align at $\approx0$\deg. Given that the emission of the hard X-ray tail is likely a combination of the quiescent and enhanced emission, it is noteworthy that the measured polarization achieves such high values. It may indicate that the locale of the outburst emission originates from a similar site as the persistent emission (likely close to the equatorial region from twisted dipolar field lines), showcasing how polarization measurements can constrain the emission topology of magnetars in outbursts. We note that the above results are fully supported in our model-independent, energy-resolved findings in which the PD increases with energy from about $15\%$ to $70\%$ in the $2-8$ keV range, while the PA remains constant at 0\deg. Lastly, phase-resolved analysis shows marginal ($2\sigma$) variation in PD with spin-phase in the $2-8$ keV range with the maximum PD aligning with the trailing shoulder of the main pulse (i.e., where the peak of the hard X-ray pulse falls), and the minimum PD is observed at pulse minimum. Our polarimetry results for \src\ display similar trends to those derived for the magnetar 1RXS J170849.0$-$400910, yet with two significant differences; in the latter (1) a soft X-ray component dominates throughout the \ixpe\ band \citep[][Stewart et al. in prep.]{zane23ApJ}, and (2) the PD displays a clear anti-correlation with the intensity pulse profile. These differences might be due to differing emission components (or emission contributions) in the $2-8$ keV range as observed in these two sources. Additionally, the fact that \src\ was observed in an enhanced state while 1RXS J1708 remained  quiescent may contribute to these differences as well.

\begin{figure*}[t!]
    \centering
    \includegraphics[width=\linewidth]{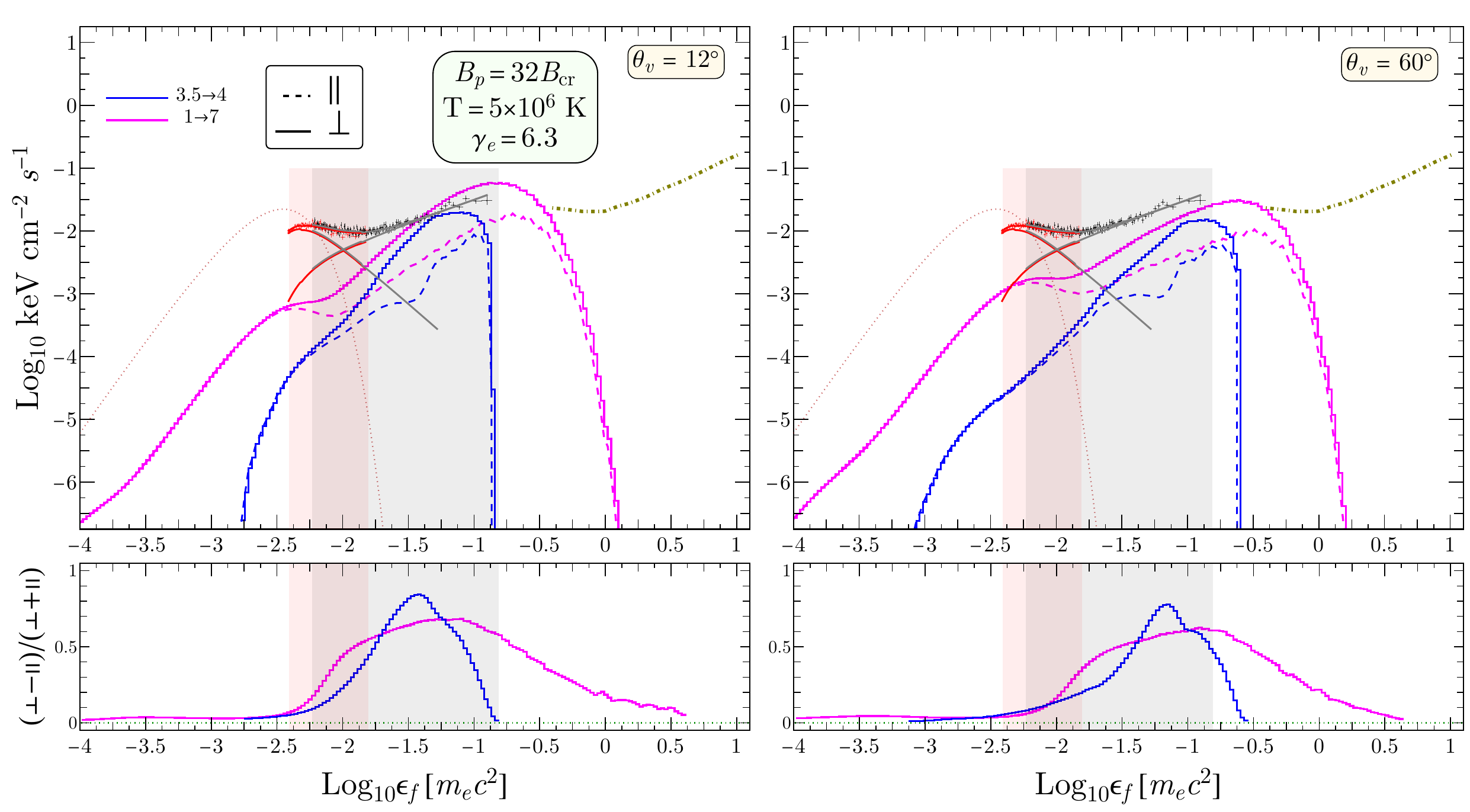}
    \caption{Pure RICS model spectra, without splitting attenuation or cascades, integrated over toroidal volumes as indicated two ranges of field footpoint altitudes in units of stellar radii,  $r_{\rm max} \in \{ 2^{3.5}, 2^{4}\}$ in blue and $r_{\rm max} \in \{ 2^1, 2^{7}\}$ in magenta. A blackbody with appropriate area and $T = 5\times 10^6$ K is plotted in dotted red. The left and right panels depict viewing angles of $\theta_v = \{12^\circ, 60^\circ\}$ with respect to the magnetic pole, as indicated. The bottom row depicts the polarization degree, which traces the $\perp$ mode preferentially resulting from scattering, which map to solid and dashed curves in the upper panel. \ixpe\ and \nustar\ data for \src\ are plotted and the energy ranges of the two instruments are indicated with relevant shading. The brown dash-dot curve depicts the 2 yr continuum sensitivity for \textit{COSI} \citep{2023arXiv230812362T}.  
    }
    \label{fig:RICS_spectra}
\end{figure*}


\subsection{Interpretation of the hard X-ray tail}

\subsubsection{Resonant Inverse Compton Scattering}

The non-thermal emission that starts dominating at energies $\gtrsim5$~keV and extends to 70~keV demands a non-thermal particle population. 
This hard X-ray component is pulsed with a broad profile and thus likely arises from low altitudes in the magnetosphere, i.e. much nearer to the rotating star than to its light cylinder.

The most efficient radiative process for generating such hard X-ray tails in magnetars  is resonant inverse Compton scattering \citep[RICS,][]{BH-2007-ApandSS,fernandez07ApJ}. Large photon densities originating from the surface bathe the inner magnetosphere, and relativistic electrons/positrons can scatter these soft X-rays to higher energies \citep{BH-2007-ApandSS,2011ApJ...733...61B}. The cross section is resonant at the cyclotron energy in the electron rest frame (ERF) \citep{Herold-1979-PhRvD,1986PhRvD..34..440B,DH-1986-ApJ}, with the scattering of photons in the $\perp$ (X-mode) state generally exceeding that of the $\parallel$ (O-mode) in head-on scatterings in the ERF. Lorentz transformations and scattering kinematics lead to a strong angle dependence of the RICS rates, and yield a profound anisotropy in emission due to Doppler beaming.  The resonance is always accessible in the thermal photon bath, and leads to a characteristic hard, flat power-law spectrum for even monoenergetic electrons \citep{2018ApJ...854...98W}.  Moreover, the resonance access condition leads to a direct mapping between observed photon energies at particular viewing directions and associated locales in the magnetosphere.  For a rotating magnetar, the emission anisotropy generates pulsation, and the pulse fraction increases with photon energy, as is observed for the non-thermal hard X-rays of several magnetars.  Yet, we note that the generation of broad pulse profiles generally requires lower Lorentz factors $\gamma_e \sim 10-30$.

\begin{figure}[t!] 
    \centering
    \includegraphics[width=\linewidth]{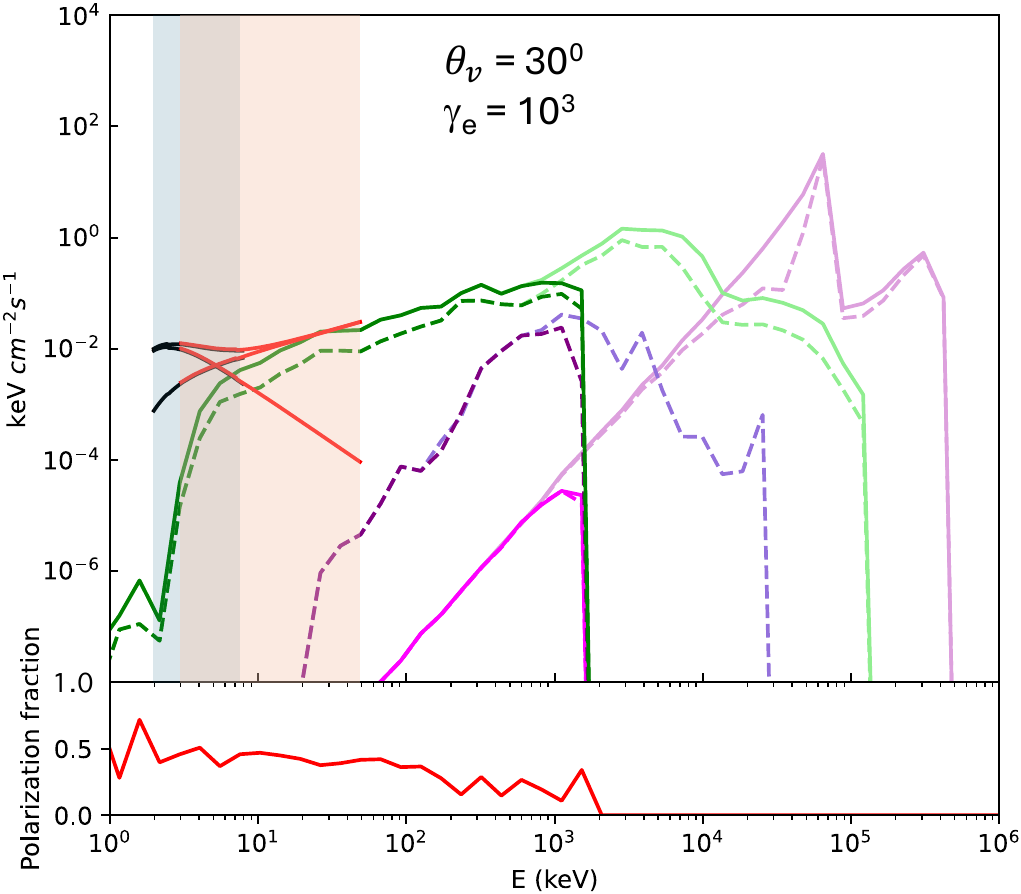}
    \caption{Top panel: Model photon spectra for O-mode (dashed lines) and X-mode (solid lines) modes from RICS (pink), splitting (purple) and pair synchrotron (green) from a pair cascade from electrons on a closed magnetic field loop with $r_{\rm max} = 6$ neutron star radii, surface magnetic field strength $B = 32\,B_{\rm cr}$, viewing angle, $\theta_v = 30^\circ$, scattering photons from the whole neutron star surface with temperature $5 \times 10^6$ K. Light lines are unattenuated spectra and dark lines are attenuated (observed) spectra.  The normalization is arbitrary.  The solid black/red lines are the soft, hard and total model fits of \ixpe/\nustar\ data.  The \ixpe\ model has been multiplied by a factor of 1.4 to correct for the calibration uncertainty with \nustar. Bottom panel: polarization degree of the total attenuated spectrum.}
    \label{fig:cascade_spectra}
\end{figure}

{\citet{2018ApJ...854...98W} predominantly considered RICS with relativistic charges confined to single field loops and monoenergetic electrons and found model spectra were too hard to match the data of persistent hard X-ray emission of the sample of quiescent magnetars. Toroidal shells of field loops steepen the spectra, yet usually not enough to accommodate magnetar observations. This motivated more recent studies \citep[first presented in][]{2019BAAS...51c.292W} which consider ensembles of relativistic charges in the magnetosphere over extended volumes.  This appears sufficient to soften pure RICS spectra for even higher particle energies $\gamma_e \sim 100$.
The accompanying generation of much more energetic photons then may lead to significant attenuation and cascading by single photon pair production and photon splitting \citep{baring01ApJ}.  Moreover, high altitude large angle scattering may result in a low-energy component below 10 keV that is largely unpolarized. The domain of pure RICS emission as a possible explanation for magnetar hard tails is therefore primarily for lower Lorentz factors where gamma-ray attenuation and pair cascading are not relevant.}

In Figure~\ref{fig:RICS_spectra} we present computations of pure RICS model spectra (Wadiasingh et al., in prep.) for a relatively low Lorentz factor of $\gamma_e \approx 6.3$ depicting various volume ranges, expressed in log base-2 maximum radius of a loop in units of stellar radius, $\log_2 r_{\rm max} $, i.e. $1\rightarrow 7$ expresses a large toroidal volume from the equator to footpoints attaining 128 stellar radii. Here $\theta_v$ is the instantaneous angle of the observer with respect to the magnetic pole, and the choice of parameters reflects the character of \src\ with $B_p \approx 32 B_{\rm cr}$, where $B_{\rm cr}  \approx 4.41 \times 10^{13}$~G is the quantum critical field, and $T \sim 5 \times 10^6$ K. One can see for these ranges of toroidal volumes, a high polarization degree of $30-70\%$ in the \ixpe\ band may be attained by pure RICS. Notable in the models is the polarization degree that is increasing with energy, similar to that observed. This selection is marginally compatible with the \nustar\ power law, although the blue curve predicts a spectral cutoff right at the edge of the \nustar\ band. Choosing more extended volumes, larger viewing angles to the magnetic axis, or higher Lorentz factors in zones of activation can result in higher energy emission, potentially extending to the \textit{COSI} band \citep{2023arXiv230812362T} although accompanied by a generally lower polarization degree in the \ixpe\ band.
\subsubsection{Synchrotron radiation from pair cascades}

The RICS radiation on closed field lines from relativistic electrons (of higher Lorentz factors than considered in Figure~\ref{fig:RICS_spectra}) with Lorentz factors $\gamma_e > 10-100$ will be attenuated by one-photon pair production and photon splitting \citep{baring01ApJ}.  In near-surface magnetar fields, $B_0 > 0.2\,B_{\rm cr}$, the pairs are produced very near the threshold, $2m_ec^2/\sin\theta_{\rm kB}$, where $\theta_{\rm kB}$ is the angle between the photon momentum and local field.  However, if the closed loops extend to altitudes of several stellar radii, the local magnetic field will drop to values that allow the pairs to be produced in higher excited states.  Photon splitting  \citep{adler70PhRvL:phSplit} does not have a threshold, so that the RICS spectra can be attenuated below pair threshold.  If the weak-dispersion regime applies in magnetar fields, then the only allowed splitting mode is the X-mode $\rightarrow $ O-mode O-mode. In this limit, photons polarized in X-mode split before reaching pair threshold and O-mode photons pair produce. The electron-positron pairs that are produced in excited Landau levels will radiate synchrotron photons of lower energy.  Those photons together with the split photons will add components to the total radiation spectrum.

Figure \ref{fig:cascade_spectra} shows the resulting spectrum at a particular viewing angle, $\theta_v = 30^\circ$, from a Monte-Carlo simulation of a pair cascade (Harding et al. in prep) initiated by electrons with Lorentz factor $\gamma_e = 10^3$ on a closed field loop with maximum radius $r_{\rm max} = 6$ R$_{\rm NS}$.  The surface magnetic field strength is $B_p = 32\,B_{\rm cr}$, to match that of \src\, and the neutron star whole-surface temperature is $5 \times 10^6$ K.  RICS, splitting and synchrotron radiation components are all visible. Since the radiation at this observer angle comes from electrons approaching the top of the field loop, the local magnetic field $B \sim 0.08\,B_{\rm cr}$ where splitting is no longer dominant and both photon polarization modes for all spectral components are attenuated by pair production. The RICS spectrum and the split photon spectrum are quite hard but with the pairs produced in high Landau states, the synchrotron spectrum overwhelms both and is much softer. The synchrotron spectrum extends from the high-energy cutoff near 2 MeV down to around 3--4 keV, with X-mode photons dominating. The polarization degree is near 50\% up to about 100--200 keV at which point the photon splitting spectrum, dominated by O-mode photons, begins to contribute. Also plotted are the \ixpe\ and \nustar\ model soft and hard power law fits as well as the total model spectrum. The hard power law model matches the pair synchrotron spectrum well for this viewing angle.  The synchrotron component is present at all viewing angles for these parameters but its spectral index varies, being softer at larger viewing angles.  Although the model shown in Figure \ref{fig:cascade_spectra} assumes that the electrons do not suffer any RICS energy losses, in cases where they do lose energy there are still dominant synchrotron components with similar photon index and polarization (Harding et al. in prep).  The PA of the radiation from closed loops will reflect the direction of the magnetic field that points to a given observer on the plane of the sky.  The uniformity of the PA across both phase and energy constrains the hard emission to locations near the tops of closed field loops, at smaller viewing angles to the magnetic pole similar to what is shown in Figure \ref{fig:cascade_spectra}, and not near the loop footpoints. This viewing direction would also allow the observer to see both the hard emission and the soft emission, if it originates near the poles.  It is likely, then, that the PA of the soft emission coming from the pole will be similar to the PA of the hard emission component from the loops.
The role of strong twists in the field structure in limiting PA variations with phase is discussed in the next Subsection.

\subsection{The nature of the soft X-ray emission}

The BB+2PL fit presented suggests that the spectrum below 5 keV is dominated by a steep PL, which is consistent with previous spectral analyses of the source \citep{an13ApJ:1841, an15ApJ:1841}.   It is appropriate to suggest a possible origin for such a soft component, to motivate deeper investigations down the line. It could be described by a Comptonization scenario, such as that applied to coronae proximate to accretion disks.  This picture was invoked in the context of bursts from the magnetar SGR J0501+4516 by \cite{Lin-2011-ApJ}.  Since the surface field is so high, the cyclotron resonance in the Compton cross section \citep[e.g.,][]{Canuto-1971-PhRvD} resides in the gamma-ray band and so the scatterings would be non-resonant.  This contrasts the repeated resonant cyclotron scattering model of \cite{Lyutikov-2006-MNRAS}, which can apply at higher altitudes of radii \teq{R\sim 20\rns} or more. Using energetics arguments supplied in \cite{BH-2007-ApandSS}, it can be easily shown that non-magnetic Thomson optical depths of \teq{\taut = n_e\sigt R  \gtrsim 1} are likely for such a luminous component. Furthermore, such densities \teq{n_e\gtrsim 3\times 10^{17}} cm$^{-3}$ can sustain current-driven magnetic field twists \citep[e.g., ][]{Thompson-2002-ApJ} of substantial magnitude, sufficient to significantly straighten field lines from the dipolar morphology in the inner magnetosphere. The strong field impacts the Comptonization process, modifying the Compton \teq{y}-parameter, and leads to a modest steepening of the spectrum \citep{Ceccobello-2014-AandA} relative to textbook non-magnetic considerations \citep[e.g.,][]{RL-book-1979}. Such a steepening reflects a modest degradation of the Comptonization efficiency by the magnetic field.  The radiation emergent from the scattering zone is beamed somewhat along the local field lines and is polarized with a dominant X mode (e.g., see Fig.~2 of \citealt{Hu-2022-ApJ}; polarization degrees in the range of 20-60\% are quite possible therein).  
Detailed modeling of the polarization expectations for such a Comptonization scenario is thus motivated, and will be the subject of future work.

As an alternative to the steep PL picture, the thermal model (2BB) that describes the soft part of the \ixpe\ spectrum is similar to the two component blackbody spectral models that are fit to the magnetars with previous \ixpe\ polarization measurements, 4U~0142+61 \citep{taverna22Sci} and 1RXS~J170849.0$-$400910 \citep{zane23ApJ}. These works argued that the surface of these two magnetars is in a condensed state, and the measured polarization degree and angle are due in part to emission from this condensed surface. On the other hand, an alternative explanation is that the polarization measurements are due to the effect of QED birefringence and mode conversion in a strong magnetized heavy element atmosphere \citep{lai23PNAS}. Unfortunately, the observations presented here for \src\ cannot discriminate between these two explanations. One reason is that the magnetic field of \src\ is stronger than that of the other two magnetars, especially that of 4U~0142+61, such that mode conversion in \src\ can produce high PD and no change in polarization angle, while condensation of part of the surface is also plausible with the inferred temperature of the large cool component \cite[][for a detailed discussion in the context of \src, see \citealt{2024arXiv241215811R}]{medin2007}. Moreover, the observations of \src\ are taken during an outburst, with possible contamination from differing components, while those of 4U~0142+61 and 1RXS~J170849.0$-$400910 occurred in quiescence.

\section*{Acknowledgments}

The authors are grateful to the \ixpe\ team for approving their DDT observation request of \src. The authors also thank the \nustar\ team for approving the DDT observation simultaneous to \ixpe\ and the \nicer\ team for continuing with the yearly magnetar monitoring program.  A portion of this work was supported by NASA through the NICER mission and the Astrophysics Explorers Program. R.S. is partly funded through NASA grants 80NSSC21K1997, 80NSSC23K1114, and 80NSSC22K0853 (PI G.Y.). The material is based upon work supported by NASA under award number 80GSFC21M0002. M.G.B. thanks NASA for generous support under awards 80NSSC22K0777, 80NSSC22K1576 and 80NSSC24K0589.  W.C.G.H. acknowledges support through grant
80NSSC23K0078 from NASA.  M. Ng is a Fonds de Recherche du Quebec – Nature et Technologies (FRQNT) postdoctoral fellow.

\bibliographystyle{aa}

\appendix

\section{Background rejection considerations}
As mentioned in the manuscript we did not perform the instrumental background rejection procedure outlined in \cite{DiMarcoCut}. In this appendix we illustrate the reason behind this choice.

\begin{figure}
    \centering
    \includegraphics[width=0.32\linewidth]{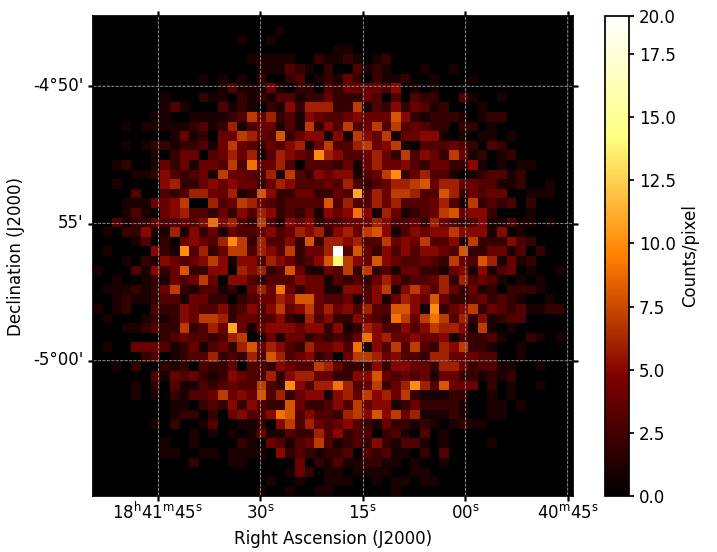}
    \includegraphics[width=0.33\linewidth]{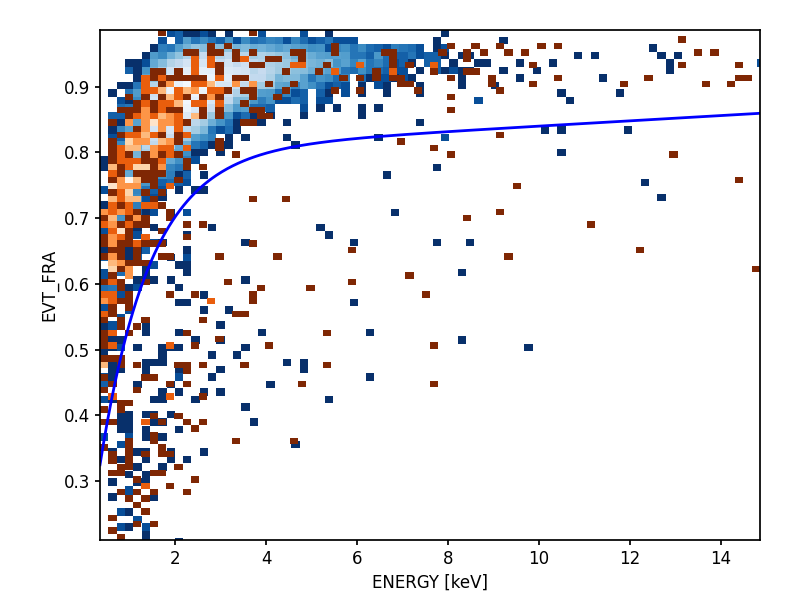}
    \includegraphics[width=0.25\linewidth]{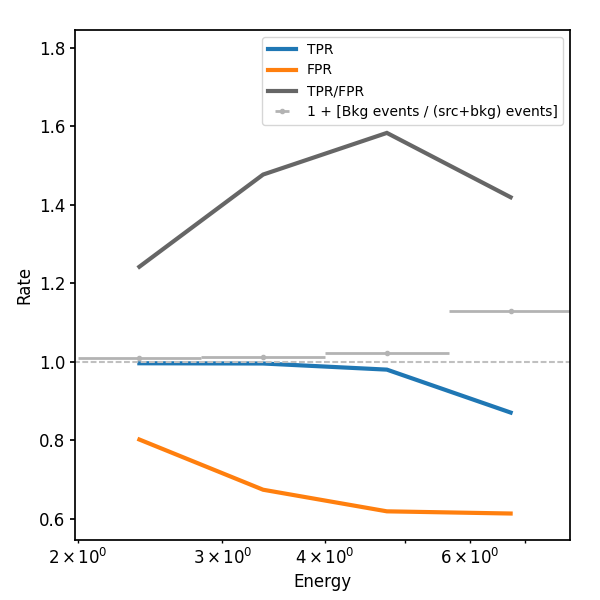}
    \caption{\textit{Left:} Counts map of the rejected component by the cut devised in \cite{DiMarcoCut}. \textit{Middle:} Heat maps (source region in blue shades, background region in red shades) of the energy fraction contained in the main cluster (EVT\_FRA) of the event reconstructed in the detector as a function of the reconstructed energy; the blue line illustrate the background rejection cut, which removes all event below it. \textit{Right:} \textit{False positive rate} (FPR) and the \textit{true positive rate} (TPR), defined as described in the text, in blue and orange respectively; the dark gray line corresponds to the ratio between the TPR and FPR; the light gray point shows the ratio between the number of counts in the background region and the source region in each energy bin (the ratio is scaled up by 1 for visualization purposes). }
    \label{fig:rejected_bkg}
\end{figure}

Fig.~\ref{fig:rejected_bkg} (left panel) shows the rejected component in the FoV resulting from the background rejection in the 6-8 keV energy band. It is evident that the cut removes good photons from the source. The middle panel illustrates the background rejection cut in blue, which reject all events below the curve defined in the space of the energy fraction contained in the main cluster (EVT\_FRA) as a function of the reconstructed energy. The blue and red maps show the number of events in that parameter space within 30 arcsec regions around the target and a background-only region (far from the SNR). This plot shows how this cut becomes more and more inefficient with increasing energy, increasingly rejecting significant fraction of source events, while not removing most of the background events. This is also shown in the plot on the right panel of Fig.~\ref{fig:rejected_bkg}. Here we illustrate the \textit{false positive rate} (FPR) and the \textit{true positive rate} (TPR) defined as follows,

\begin{equation}
    FPR = \frac{FP}{FP+TN} ~~~~~~
    TPR = \frac{TP}{TP+FN},
\end{equation}

where $FP$ is the number of events in the background-only region we keep after the cut (false positives), $TN$ is the number of events in the background-only region we reject with the cut (true negatives), $TP$ is the number of events in the source region we keep after the cut (true positives), and $FN$ is the number of events in the source region we reject with the cut (false negatives). Ideally one would like to maximize the TPR (the efficiency in keeping source events) while minimizing the FPR (the background contamination). It is evident that above 5 keV the cut becomes inefficient, lowering the true positive rate while not improving on the false positive rate. As mentioned in the text, below 5 keV the count statistics of the source is high, dominating over the instrumental background. We hence conclude that the background rejection cut is only beneficial but ineffective below 5 keV, while becoming increasingly counterproductive at higher energies, ultimately failing to improve the signal-to-background ratio. This led to the decision to not apply the background rejection procedure for the analysis of this \ixpe\ observation, relying on the subtraction of the unpolarized SNR as background as a sufficient method to remove both instrumental and astrophysical background contamination (see Section \ref{sec:obs}). 

\end{document}

We additionally fit a normalization constant for each detector to track the cross-calibration factors between the telescopes. We find the maximum gap to be $\sim 40\%$ between \textit{IXPE} DU3 and \nustar~ FPMA. Along with the cross-calibration uncertainties between \nustar~ and \textit{IXPE}'s detectors, the discrepancy can likely be attributed to \textit{IXPE}'s superior ability to resolve Kes 73 which makes consistent treatment of the low-energy contributions from the SNR across instruments difficult. There additionally appears to be a marginal discrepancy between the two \textit{IXPE} observations of \src\ (see Appendix for details of the time-resolved spectroscopy). These combined effects may account for the observed $40\%$ disparity, though further investigation may be warranted.